\documentstyle[12pt]{article}
\def\hybrid{\topmargin 0pt      \oddsidemargin 0pt
        \headheight 0pt \headsep 0pt
        \textwidth 17.5cm
        \textheight 25cm
        \voffset=-0.7cm
        \hoffset=-0.4cm
       \hoffset=-1.2cm
        \marginparwidth 0.0in
        \parskip 5pt plus 1pt   \jot = 1.5ex}
\catcode`\@=11
\def\marginnote#1{}

\newcount\hour
\newcount\minute
\newtoks\amorpm
\hour=\time\divide\hour by60
\minute=\time{\multiply\hour by60 \global\advance\minute by-\hour}
\edef\standardtime{{\ifnum\hour<12 \global\amorpm={am}%
        \else\global\amorpm={pm}\advance\hour by-12 \fi
        \ifnum\hour=0 \hour=12 \fi
        \number\hour:\ifnum\minute<10 0\fi\number\minute\the\amorpm}}
\edef\militarytime{\number\hour:\ifnum\minute<10 0\fi\number\minute}

\def\draftlabel#1{{\@bsphack\if@filesw {\let\thepage\relax
   \xdef\@gtempa{\write\@auxout{\string
      \newlabel{#1}{{\@currentlabel}{\thepage}}}}}\@gtempa
   \if@nobreak \ifvmode\nobreak\fi\fi\fi\@esphack}
        \gdef\@eqnlabel{#1}}
\def\@eqnlabel{}
\def\@vacuum{}
\def\draftmarginnote#1{\marginpar{\raggedright\scriptsize\tt#1}}

\def\draft{\oddsidemargin -0.1truein
        \def\@oddfoot{\sl preliminary draft \hfil
        \rm\thepage\hfil\sl\today\quad\militarytime}
        \let\@evenfoot\@oddfoot \overfullrule 3pt
        \let\label=\draftlabel
        \let\marginnote=\draftmarginnote
   \def\@eqnnum{{\rm (\theequation)}\rlap{\kern\marginparsep\tt\@eqnlabel}%
\global\let\@eqnlabel\@vacuum}  }
\newdimen\linethick  \linethick=0.4pt
\newdimen\hboxitspace    \hboxitspace=5pt
\newdimen\vboxitspace    \vboxitspace=5pt

\def\fr#1{%
\beq\new
\vcenter{
\hrule height\linethick
           \hbox{\vrule width\linethick
                 \kern\hboxitspace
                 \vbox{\kern\vboxitspace
                       \hbox{$\begin{array}{c}\displaystyle#1
          \end{array}$}%
                       \kern\vboxitspace}%
                 \kern\hboxitspace
                 \vrule width\linethick}%
           \hrule height\linethick}%
\eeq}
\newdimen\Squaresize \Squaresize=14pt
\newdimen\Thickness \Thickness=0.5pt

\def\Square#1{\hbox{\vrule width \Thickness
   \vbox to \Squaresize{\hrule height \Thickness\vss
      \hbox to \Squaresize{\hss#1\hss}
   \vss\hrule height\Thickness}
\unskip\vrule width \Thickness}
\kern-\Thickness}

\def\Vsquare#1{\vbox{\Square{$#1$}}\kern-\Thickness}

\def\numberbysection{\@addtoreset{equation}{section}
        \def\theequation{\thesection.\arabic{equation}}}
\numberbysection

\renewcommand{\theequation}{\thesection.\arabic{equation}}
\newcommand{\l@qq}[2]{\addvspace{2em}
 \hbox to\textwidth{\hspace{1em}\bf #1 \dotfill #2}}

\newcounter{app}

\def\app{\setcounter{equation}{0}
\def\theequation{\Alph{app}.\arabic{equation}}\par
   \addvspace{4ex}
   \@afterindentfalse
  \secdef\@app\@dapp}
\newcommand\@app{\@startsection {app}{1}{0ex}%
                                   {-3.5ex \@plus -1ex \@minus -.2ex}%
                                   {2.3ex \@plus.2ex}%
                                   {\normalfont\Large\bf}}

\def\@dapp#1{%
{\parindent \z@ \raggedright  \bf #1}\par\nobreak}
\def\l@app#1#2{\ifnum \c@tocdepth >\z@
    \addpenalty\@secpenalty
    \addvspace{1.0em \@plus\p@}%
    \setlength\@tempdima{2.5em}%
    \begingroup
      \parindent \z@ \rightskip \@pnumwidth
      \parfillskip -\@pnumwidth
      \leavevmode \bfseries
      \advance\leftskip\@tempdima
      \hskip -\leftskip
      #1\nobreak\hfil \nobreak\hb@xt@\@pnumwidth{\hss #2}\par
    \endgroup\fi}
\newcounter{sapp}[app]

\def\sapp{\def\theequation{\Alph{app}.\arabic{equation}}\par
   \@afterindentfalse
  \secdef\@sapp\@dsapp}
\newcommand\@sapp{\@startsection{sapp}{2}{\z@}%
                                     {-3.25ex\@plus -1ex \@minus -.2ex}%
                                     {1.5ex \@plus .2ex}%
                                     {\normalfont\large\bfseries}}

\def\@dsapp#1{%
{\parindent \z@ \raggedright  \bf #1}\par\nobreak}
\newcommand{\l@sapp}{\@dottedtocline{2}{1.5em}{3em}}

\def\titlepage{\@restonecolfalse\if@twocolumn\@restonecoltrue\onecolumn
     \else \newpage \fi \thispagestyle{empty}\c@page\z@
        \def\thefootnote{\fnsymbol{footnote}} }

\def\endtitlepage{\if@restonecol\twocolumn \else  \fi
        \def\thefootnote{\arabic{footnote}}
        \setcounter{footnote}{0}}  %\c@footnote\z@ }
\relax

\hybrid

\newtoks\@stequation

\def\subequations{\refstepcounter{equation}%
  \edef\@savedequation{\the\c@equation}%
  \@stequation=\expandafter{\theequation}%   %only want \theequation
  \edef\@savedtheequation{\the\@stequation}% %expanded once
  \edef\oldtheequation{\theequation}%
  \setcounter{equation}{0}%
  \def\theequation{\oldtheequation\alph{equation}}}

\def\endsubequations{%
  \setcounter{equation}{\@savedequation}%
  \@stequation=\expandafter{\@savedtheequation}%
  \edef\theequation{\the\@stequation}%
  \global\@ignoretrue}

\makeatletter
\newdimen\normalarrayskip              % skip between lines
\newdimen\minarrayskip                 % minimal skip between lines
\normalarrayskip\baselineskip
\minarrayskip\jot
\newif\ifold             \oldtrue            \def\new{\oldfalse}
\def\arraymode{\ifold\relax\else\displaystyle\fi} % mode of array enrties
\def\eqnumphantom{\phantom{(\theequation)}}     % right phantom in eqnarray
\def\@arrayskip{\ifold\baselineskip\z@\lineskip\z@
     \else
     \baselineskip\minarrayskip\lineskip1\baselineskip\fi}

\def\@arrayclassz{\ifcase \@lastchclass \@acolampacol \or
\@ampacol \or \or \or \@addamp \or
   \@acolampacol \or \@firstampfalse \@acol \fi
\edef\@preamble{\@preamble
  \ifcase \@chnum
     \hfil$\relax\arraymode\@sharp$\hfil
     \or $\relax\arraymode\@sharp$\hfil
     \or \hfil$\relax\arraymode\@sharp$\fi}}

\def\@array[#1]#2{\setbox\@arstrutbox=\hbox{\vrule
     height\arraystretch \ht\strutbox
     depth\arraystretch \dp\strutbox
     width\z@}\@mkpream{#2}\edef\@preamble{\halign \noexpand\@halignto
\bgroup \tabskip\z@ \@arstrut \@preamble \tabskip\z@ \cr}%
\let\@startpbox\@@startpbox \let\@endpbox\@@endpbox
  \if #1t\vtop \else \if#1b\vbox \else \vcenter \fi\fi
  \bgroup \let\par\relax
  \let\@sharp##\let\protect\relax
  \@arrayskip\@preamble}
\def\eqnarray{\stepcounter{equation}%
              \let\@currentlabel=\theequation
              \global\@eqnswtrue
              \global\@eqcnt\z@
              \tabskip\@centering                      %formulae centering
              \let\\=\@eqncr
              $$%
            \halign to \displaywidth  \bgroup
             \eqnumphantom \@eqnsel
      \hskip\@centering                               %right tab%
    $\displaystyle  \tabskip\z@ {##}$%
    &\global\@eqcnt\@ne \hskip 2\arraycolsep
         $ \displaystyle  \arraymode{##}$\hfil
    &\global\@eqcnt\tw@ \hskip 2\arraycolsep
         $\displaystyle\tabskip\z@{##}$\hfil
         \tabskip\@centering
    &{##}\tabskip\z@\cr}
\makeatother
\def\bea{\begin{eqnarray}}
\def\eea{\end{eqnarray}}

\def\beq{\begin{equation}}
\def\eeq{\end{equation}}
\def\be{\beq\new\begin{array}{c}}
\def\ee{\end{array}\eeq}
\def\bse{\begin{subequations}}                %%%SUBEQUATIONS
\def\ese{\end{subequations}}                 %
\begin{document}
\vspace{0.2cm}
\begin{center}
{\LARGE \bf Confinement versus asymptotic freedom} \\
\vspace{0.7cm} {\large\bf Andrey  Dubin}\\
\vspace{0.5cm}{\bf ITEP, B.Cheremushkinskaya 25, Moscow 117259, Russia}\\
\end{center}
\vspace{0.3cm}

\begin{abstract}

We put forward the low-energy confining asymptote of the solution
$<W_{C}>$ (valid for large macroscopic contours $C$ of the size
$>>1/\Lambda_{QCD}$) to the large $N$ Loop equation in the $D=4$ $U(N)$
Yang-Mills theory with the asymptotic freedom in the ultraviolet domain.
Adapting the multiscale decomposition characteristic of the Wilsonean
renormgroup, the proposed Ansatz for the loop-average is composed in order to
sew, along the lines of the bootstrap approach, the large $N$ weak-coupling
series for high-momentum modes with the
$N\rightarrow{\infty}$ limit of the recently suggested stringy representation
of the $1/N$ strong-coupling expansion \cite{Dub4} applied to low-momentum
excitations. The resulting low-energy stringy theory can be described through
such superrenormalizable deformation of the noncritical Liouville string that,
being devoid of ultraviolet divergences, does not possess propagating degrees
of freedom at short-distance scales $<<1/{\sqrt{\sigma_{ph}}}$, where
$\sigma_{ph}\sim{(\Lambda_{QCD})^{2}}$ is the physical string tension.

\end{abstract}                                                

\begin{center}
\vspace{0.5cm}{Keywords: Yang-Mills, Loop equation, String,
Strong-coupling expansion} \\
\vspace{0.2cm}{PACS codes 11.15.Pg; 11.15.Me; 12.38.Aw; 12.38.Lg}
\end{center}

\newpage

\section{Introduction}

The determination of the mechanism of confinement in the asymptotically free
$4$-dimensional $SU(N)$ theory, defined by the action
\be
S=\frac{N}{4\lambda}\int d^{4}x~tr\left( F_{\mu\nu}(x)F_{\mu\nu}(x)\right)~,
\label{1.1}
\ee
is one of the fundamental problems of the contemporary physics.
Despite that the latter theory was proposed long ago, we still possess very
limited and mostly qualitative understanding of the latter phenomenon, and
the Loop equation \cite{LE/MM,LE/P} remains to be one of the major tools
which can be utilized to attack this problem. Recently, these equations
has been utilized in \cite{Dub4} to show that confinement holds true, at
least for $N$ large enough, in the $3\leq{D}\leq{D_{cr}}$ option
(with $D_{cr}>4$) of the regularized $U(N)$ or
$SU(N)$ Yang-Mills theory (\ref{1.1}) considered in the unconventional
strong-coupling phase, without asymptotic freedom, when in eq. (\ref{1.1})
the {\it bare} coupling constant $\lambda$ is sufficiently large. In this
phase, the adequate expansion for the averages of the Wilson loops
\be
W_{C}=\frac{1}{N}
Tr\Bigg[~{\mathcal{P}}~exp\left(i\oint_{C} dx_{\mu}(s)A_{\mu}(x)
\right)\Bigg]~,
\label{1.2}
\ee
is provided by the $1/N$ {\it strong-coupling}\footnote{Owing to the formal
resummation \cite{Dub3} between the $1/N$ $WC$ and $1/N$ $SC$ expansions in a
generic pure gauge model, the considered $SC$ series can be reinterpreted as
the higher-dimensional generalization of the Gross-Taylor representation
\cite{Gr&Tayl} for the $1/N$ $SC$ series in the $D=2$ Yang-Mills theory.}
($SC$), rather than weak-coupling ($WC$), series. Consequently, in parallel
with the dynamics of the lattice gauge theories in the $SC$ phase, confinement
arises as the straighforward consequence of the fact that the excitations,
pertinent for the considered $1/N$ $SC$ expansion, are strings rather than
point-like particles.

In the present paper, we
propose how the results of \cite{Dub4} can be adapted to the $D=4$
$U(\infty)\cong{SU(\infty)}$ gauge system (\ref{1.1}) in the conventional
large $N$ regime when $\lambda$ vanishes,
\be
\lambda~\longrightarrow{~\frac{24\pi^{2}}{11}
\frac{1}{log(\Lambda/\Lambda_{QCD})}}~\longrightarrow~{0}~,
\label{9.11}
\ee
according to the asymptotical prediction of the large $N$ $WC$ expansion,
where $\Lambda$ denotes the $UV$ cut off sent to infinity in the units of
$\Lambda_{QCD}$. Adapting the multiscale decomposition \cite{MSD} routed in the Wilsonean
approach to the renormgroup, in the limit (\ref{9.11}) we propose the
low-energy asymptote of the solution of the $U(\infty)$ Loop
equation in the form
\be                                                      
<W_{C}>_{\infty}=<W_{C}>_{1}<W_{C}>_{2}\Big|_{N\rightarrow{\infty}}
\label{9.2}
\ee
composed of the two factors $<W_{C}>_{k}$ associated to a given macroscopic
contour $C$ possessing the radius of curvature ${\cal R}(s)>>1/\Lambda_{QCD}$.
The part $<W_{C}>_{1}$, accumulating the contribution (\ref{9.30})
from short distances $<b\Lambda_{QCD}^{-1}$ (where $b$ is some 
constant specified further below), is to be reproduced via
the large $N$ $WC$ series. Compared to the standard perturbative approach,
this series are {\it modified} in the infrared domain
due to certain gauge-invariant infrared cut off at the scale
$\sim{b\Lambda_{QCD}^{-1}}$. As for the factor $<W_{C}>_{2}$, being associated
to the low-energy excitations, it accumulates the complementary contribution
into $<W_{C}>_{\infty}$ that refers to the infrared domain of distances
$>b\Lambda_{QCD}^{-1}$ where the gauge theory (\ref{1.1}) is supposed to be
strongly coupled. Therefore, the analysis \cite{Dub4} of the $SC$ phase
suggests to describe $<W_{C}>_{2}$ through the
$N\rightarrow{\infty}$ limit of the low-energy asymptote of the $1/N$ $SC$
expansion. The latter asymptote, being presumably universal for a generic
pure gauge model, is represented by the judicious implementation of the
large $N$ stringy Ansatz \cite{Dub4} predetermined by the
concise worldsheet's weight\footnote{The weight (\ref{0.1}) belongs to the
more general variety of confining strings \cite{PolyakCS} suggested by
A.Polyakov as possible candidates for the solution of the Loop equation.}
\be
\bar{w}_{2}[\tilde{M}]=
exp\left(-\frac{\lambda{\cal M}^{2}}{4}
\int\limits_{\tilde{M}}\int\limits_{\tilde{M}}
d\sigma_{\mu\nu}({\bf x})d\sigma_{\mu\nu}({\bf y})~
{\cal M}^{2}{\cal F}_{2}({\cal M}^{2}({\bf x}-{\bf y})^{2})
\right)~,
\label{0.1}          
\ee
where $d\sigma_{\mu\nu}({\bf x})$ is the standard infinitesimal
area-element associated to the {\it genus-zero} connected surface
$\tilde{M}\equiv{\tilde{M}(C)}$, and the parameter
${\cal M}\sim{\Lambda_{QCD}}$ is chosen (building on the lattice
computations) to be close to the lowest glueball mass.
As for the function ${\Lambda}^{4}
{\cal F}_{2}({\cal M}^{2}{\bf z}^{2})$, being introduced
to smear the $4$-dimensional $\delta_{4}({\bf z})$-function at the scales
$\leq{\cal M}^{-1}$, it is further
constrained by the conditions (\ref{9.18}) and (\ref{9.27}) discussed below.
Complementary, for the contours without zigzag backtrackings, the averaging
over the worldsheets $\tilde{M}$ is performed through the standard
Nambu-Goto measure endowed with the $UV$ cut off (for the worldsheet's
fluctuations) of order of the inverse width $\sqrt{<{\bf r}^{2}>}
\sim{\cal M}^{-1}$ of the vortex described by the bilocal weight (\ref{0.1}).
As in \cite{Dub4}, the construction is completed
by the prescription (sketched in Section 4) to implement the
mandatory backtracking invariance of the loop-averages $<W_{C}>$
which is lacking in the conventional Nambu-Goto paradigm.

Actually, for thus defined decomposition (\ref{9.2}) to be consistent with the
Loop equation, the applicability of the stringy system (\ref{0.1}) is
restricted by the conditions similar to the ones \cite{Dub4} formulated in the
context of the $SC$ phase of the gauge theory (\ref{1.1}). Selecting the
appropriate variety of the functions ${\cal F}_{2}$, one is to focus on
certain class $\Upsilon'$ of macroscopic contours
$C$ (of the radius of curvature ${\cal R}(s)>>{\cal M}^{-1}$) so that the
considered stringy system approaches, from the viewpoint of the Wilsonean
renormgroup, its low-energy asymptote. For the contours $C$ devoid of zigzag
backtrackings, this asymptote fits in the pattern provided by the
low-energy limit of the simpler system: the unconventional implementation
(owing to the condition (\ref{9.18}) below) of the good old Nambu-Goto theory
with the properly identified parameters. The latter theory, being presumed to
possess the same {\it UV} cut off ${\cal M}$ as the system (\ref{0.1}) does,
is endowed with the weight
\be                            
\bar{w}_{1}[\tilde{M}]=
exp\left({-\frac{\bar{\lambda}{\cal M}^{2}}{2}\cdot
A[\tilde{M}]}-\bar{\lambda}_{1}{\cal M}\cdot L[\partial\tilde{M}]\right)~,
\label{2.5bb}
\ee
where $\bar{\lambda}$ and $\bar{\lambda}_{1}$ are the auxiliary dimensionless
parameters, while $A[\tilde{M}]$ and $L[\partial\tilde{M}]$ denote
respectively the total area and the length of the boundary
$\partial{\tilde{M}}=C$ both associated to a given genus-zero worldsheet
$\tilde{M}$. Secondly, the stringy system endowed with the weight
(\ref{2.5bb}) should confine with the string tension fulfilling the
specific ${\cal M}^{2}$-scaling
\be
\sigma_{ph}~\sim{~{\cal M}^{2}}~,
\label{9.18}
\ee
which implies that $\sqrt{\sigma_{ph}}$ is {\it of order of}, rather than much
smaller or much larger than, the $UV$ cut off for the worldsheet's
fluctuations. Being
unable at present to determine the precise value of $\sigma_{ph}$, the scaling
(\ref{9.18}) is as far as we can reach with the help of such implementation
(based on eq. (\ref{0.1})) of $<W_{C}>_{2}$ that is inapplicable
in the case of the contours of the size $<<{1/\Lambda_{QCD}}$.

In Section 2, we sketch the regularization of the Loop equation which,
building on the gauge-invariant prescription of \cite{MigdRep,MakRev},
provides with the appropriate starting point of our analysis.
The counterpart of the field-theoretic variant of the multiscale
decomposition \cite{MSD}, suitable in the context of the Loop equation,
is introduced in Section 3. In Sections 4 and 5, the asserted pattern of the
Ansatz (\ref{9.2}) is derived utilizing the bootstrap approach to sew the
large $N$ $WC$ and $SC$ series. In Conclusions, we
suggest how the unconventional implementation (\ref{9.18}) of the Nambu-Goto
theory (\ref{2.5bb}) can be reproduced in the paradigm of certain
superrenormalizable deformation of the noncritical Liouville string.
The Appendix outlines the mechanism that ensures the infrared
equivalence between the stringy systems endowed respectively with the weights
(\ref{0.1}) and (\ref{2.5bb}).

\section{The appropriate regularization of the Loop equation}

The general pattern \cite{MigdRep,MakRev} of the gauge-invariant
regularization, see eq. (\ref{0.9zze}) below, is known to render the
Loop equation rather heavy for the brute force attack. The better alternative
\cite{Dub4} is to take advantage of the specific simplification that arises
for the macroscopic contours $C$ (parameterized by the trajectory
${\bf x}(s)$) which, possessing the radius of curvature
${\cal R}(s)>>1/\cal M$, belong to the subspace $\Upsilon'$ further specified
shortly below. In this case, one can introduce the two somewhat
distinct regularization prescriptions that, being consistent with each other,
are associated prior to the regularization respectively to the contribution of
the trivial (i.e., when ${\bf x}(s)={\bf y}(s')$ with $s=s'$) and
nontrivial (i.e., when ${\bf x}(s)={\bf y}(s')$ with $s\neq{s'}$) point-like
self-intersections of $C\in{\Upsilon'}$
into the r.h. side of the Loop equation. After the regularization, we are to
differentiate between those 'irregularities' of the contour's geometry
which, from the macroscopic viewpoint\footnote{From this viewpoint, one does
not distinguish between the points of $C$ separated by a distance
$\leq{\cal M}^{-1}$.}, as well can be viewed as point-like self-intersections. On the other hand, the relevant
for our further analysis subspace $\Upsilon'$ of the loop space is postulated
to exclude those geometries of $C$ which, from the macroscopic viewpoint, can
be viewed as one-dimensional self-intersections. For convenience, in what
follows the admissible irregularities of $C\in{\Upsilon'}$ (definition of
which is formalized in Appendix \ref{macr}) are called
quasi-self-intersections.

Concerning the case when the contour $C\in{\Upsilon'}$ is devoid of 
nontrivial quasi-self-intersections, one is to utilize the
{\it linearization} of the nonregularized Loop equation \cite{LE/MM,LE/P} on
the subspace of non-self-intersecting contours. (With our conventions, each
point ${\bf x}(s)$ of a contour $C$ should be interpreted as the point of
trivial self-intersection.) For such contours, it suggests the simple
regularization prescription that results in the reduced equation \cite{Dub4}
\be
\hat{\mathcal{L}}<W_{C}>_{\infty}=\lambda
\oint_{C} dx^{\nu}(s)\oint_{C} dy_{\nu}(s')
<{\bf x}(s)|~\hat{\cal G}_{\Lambda}|~{\bf y}(s')>
<W_{C}>_{\infty}~,
\label{0.9zax}
\ee
where the ${\Lambda}^{4}{\mathcal{G}}({\Lambda}^{2}{\bf z}^{2})$, being
alternatively represented via the corresponding operator
$\hat{\cal G}_{\Lambda}$, denotes a sufficiently smooth function that
maintains the smearing
\be
\delta_{4}({\bf x}-{\bf y})~\longrightarrow{~
{\Lambda}^{4}{\mathcal{G}}({\Lambda}^{2}({\bf x}-{\bf y})^{2})}
\equiv{<{\bf x}~|~\hat{\cal G}_{\Lambda}|~{\bf y}>}
~~~~~,~~~~~\int d^{4}z~{\mathcal{G}}({\bf z}^{2})=1~,
\label{0.1bb}
\ee
of the $4$-dimensional delta-function at the scale $\Lambda$ of the $UV$
regularization of the gauge theory (\ref{1.1}). As for the
{\it first} order Loop operator
\be
\hat{\cal L}=~\oint_{C} dx^{\nu}(s)
\hat{\cal L}_{\nu}({\bf x}(s))~~~~~~,~~~~~~~
\hat{\cal L}_{\nu}({\bf x}(s))=
\partial_{\mu}^{{\bf x}(s)}~
\frac{\delta}{\delta \sigma_{\mu\nu}({\bf x}(s))}~,
\label{1.6bf}
\ee
it is composed, see \cite{MigdRep} for the details, 
of the Mandelstam area-derivative $\delta/\delta\sigma_{\mu\nu}({\bf x}(s))$
and the path-derivative $\partial_{\mu}^{{\bf x}(s)}$.

Eq. (\ref{0.9zax}) can not be applied to properly handle the contribution of
the nontrivial (quasi-)self-intersections of $C\in{\Upsilon'}$. To deal with
the latter contribution, we employ the gauge-invariant regularization
\cite{MigdRep,MakRev} that allows to write down the Loop equation in the form
$$
\hat{\mathcal{L}}<W_{C}>_{\infty}~=\lambda
\oint\limits_{C} dx^{\nu}\oint\limits_{C} dy_{\nu}
<{\bf x}|\hat{\cal X}_{\Lambda}|{\bf y}>\times
$$
\be
\times\int\limits_{\Gamma_{xy}\in{{\bf X}^{\tilde{D}}}}
{\mathcal{D}}z_{\mu}(t)
<W_{C_{xy}\Gamma_{yx}}>_{\infty}<W_{C_{yx}\Gamma_{xy}}>_{\infty}
e^{-S(\Gamma_{xy})}
\label{0.9zze}
\ee
where $<{\bf x}|\hat{\cal X}_{\Lambda}|{\bf y}>=
{\Lambda}^{4}{\cal X}({\Lambda}{\bf z})|_{{\bf z}={\bf x}-{\bf y}}$ is some
regularization-dependent function introduced for the later convenience, and
in the r.h. side of eq. (\ref{0.9zze}) the functional integral runs over the
auxiliary (generically non-self-intersecting) paths $\Gamma_{xy}$ endowed with
an effective action $S(\Gamma_{xy})$ to be fixed in due course. Being
parameterized by the trajectory $z_{\mu}(t)$ (with ${\bf z}(0)={\bf x}$ and
${\bf z}(1)={\bf y}$) embedded into some $\tilde{D}\leq{4}$ dimensional domain
${\bf X}^{\tilde{D}}$ of the $4$-dimensional Euclidean space ${\bf R}^{4}$,
the paths $\Gamma_{xy}=\Gamma^{-1}_{yx}$ refer to the decomposition of the
loop $C\equiv{C_{xx}}=C_{xy}C_{yx}$ into the two (open, when
${\bf x}(s)\neq{{\bf y}(s')}$) paths $C_{xy}$ and $C_{yx}$
which, in turn, are associated to the auxiliary closed loops
$C_{xy}\Gamma_{yx}$ and $C_{yx}\Gamma_{xy}$ respectively.
Complementary, one is to require that the combination
\be
<{\bf x}~|~\hat{\mathcal{E}}_{\Lambda}|~{\bf y}>=
<{\bf x}~|~\hat{\mathcal{X}}_{\Lambda}|~{\bf y}>
\int\limits_{{\bf X}^{\tilde{D}}} {\mathcal{D}}z_{\mu}(t)~e^{-S(\Gamma_{xy})}
~~~~~~,~~~~~~
\int d^{4}z <{\bf z}~|~\hat{\mathcal{E}}_{\Lambda}|~0>=1~,
\label{1.3zaz}
\ee
provides, akin to eq. (\ref{0.1bb}), with the smearing of
$\delta_{4}({\bf z})$ at the scale $\Lambda$. In particular, in eq.
(\ref{1.3zaz}) the effective action $S(\Gamma_{xy})$ is  constrained by the
natural condition that, once $|{\bf x}(s)-{\bf y}(s')|\leq{\Lambda}^{-1}$, the
characteristic length $<L[\Gamma_{xy}]>$ of $\Gamma_{xy}$ is of order of the
inverse $UV$ cut off $\Lambda$. On the other hand, in the formal limit
$<{\bf z}~|~\hat{\mathcal{E}}_{\Lambda}|~0>\rightarrow{\delta_{4}({\bf z})}$,
eq. (\ref{0.9zze}) evidently approaches the nonregularized form
\cite{LE/MM,LE/P} of the Loop equation the r.h. side of which is formulated
in terms of the product $<W_{C_{xy}}>_{\infty}<W_{C_{yx}}>_{\infty}$.

\section{The multiscale decomposition of the Loop equation}

To make the Ansatz (\ref{9.2}) consistent with thus regularized $D=4$ Loop
equation, we propose to adapt the general technique of the multiscale
decomposition \cite{MSD}. Akin to the previous section, to develop the
appropriate formalism, it is convenient to introduce the two distinct
implementations of the decomposition associated respectively to eqs.
(\ref{0.9zax}) and (\ref{0.9zze}) corresponding to trivial and
nontrivial quasi-self-intersections.

\subsection{The trivial quasi-self-intersections}

For macroscopic contours $C\in{\Upsilon'}$ without nontrivial
quasi-self-intersections, the approach \cite{MSD} motivates to begin with
the following decomposition
\be
\hat{\cal G}_{\Lambda}=\hat{\cal G}_{1}+\hat{\cal G}_{2}~~~~~~,~~~~~~~
<{\bf z}~|~\hat{\cal G}_{k}|~0>\equiv{
{\Lambda}^{4}\cdot{\mathcal{G}}_{k}({\Lambda}^{2}{\bf z}^{2})}~,
\label{9.6a}
\ee
of the operator $\hat{\cal G}_{\Lambda}$ entering eq. (\ref{0.9zax}).
Then, combining eqs. (\ref{9.2}) and (\ref{9.6a}) and employing the fact that
the first order Loop operator (\ref{1.6bf}) complies with the Leibnitz rule
$\hat{\mathcal{L}}\left(<W_{C}>_{1}<W_{C}>_{2}\right)=\left(\hat{\mathcal{L}}
<W_{C}>_{1}\right)<W_{C}>_{2}+
<W_{C}>_{1}\left(\hat{\mathcal{L}}<W_{C}>_{2}\right)$,
we can transform eq. (\ref{0.9zax}) into the pair of the decoupled
equations, 
\be
\hat{\mathcal{L}}<W_{C}>_{k}=\lambda 
\oint\limits_{C} dx^{\nu}(s)\oint\limits_{C} dy_{\nu}(s')
<{\bf x}(s)|~\hat{\cal G}_{k}~|~{\bf y}(s')><W_{C}>_{k}~,
\label{0.9zaz}
\ee
to be labeled by index $k=1,2$, and we omit the subscript
$\infty$ of the loop-averages from now on.

As for the hierarchy of scales required for the consistency of the
decomposition, it is to be  maintained imposing the
condition that the $4$-dimensional Fourier image $\tilde{\cal G}_{k}({\bf p})$
of ${\cal G}_{k}({\bf z})=<{\bf z}|\hat{\cal K}_{k}|0>$ vanishes sufficiently
fast (compared to $\tilde{\cal G}_{\Lambda}({\bf p})$) for those ${\bf p}^{2}$
which are either $<<{\cal M}^{2}$ for $k=1$ or $>>{\cal M}^{2}$ for $k=2$.
The simplest way to formalize the considered hierarchy is to
constrain the decomposition (\ref{9.6a}) by the two conditions
\be
\int d^{4}z~{\mathcal{G}}_{2}({\bf z}^{2})=
\left({\cal M}/\Lambda\right)^{4}~~~~~~~,~~~~~~~~
<{\bf r}^{2}>_{{\cal G}_{2}}~\sim{~\cal M}^{-2}~,
\label{9.8}
\ee
where $\sqrt{<{\bf r}^{2}>}$ is the characteristic width of the vortex
that is described by the stringy representation of $<W_{C}>_{2}$ associated
to the weight (\ref{0.1}). In turn, advancing ahead, let us note that the
dependence of the latter weight on the function
${\mathcal{G}}_{2}({\bf z}^{2})$ is fixed by the relation
\be
{\cal M}^{4}\cdot{\cal F}_{2}({\cal M}^{2}{\bf z}^{2})
\equiv{{\Lambda}^{4}\cdot{\cal G}_{2}({\Lambda}^{2}{\bf z}^{2})}
\label{9.9}
\ee
which introduces the properly rescaled counterpart ${\cal F}_{2}(..)$ of
${\cal G}_{2}(..)$. As a result, the width in eq. (\ref{9.8}) is to be
naturally estimated as
\be
\sqrt{<{\bf r}^{2}>}~\sim{~{\cal M}^{-1}\cdot
supr\left(\sqrt[n]{<K_{n}/K_{0}>}~\right)}~~~~~~,~~~~~~
K_{n}[{\mathcal{F}}_{2}]=
\int_{{\mathcal{P}}} d^{2}z~({\bf z}^{2})^{\frac{n}{2}}
{\mathcal{F}}_{2}({\bf z}^{2})~,
\label{0.3csb}
\ee
where $supr(..)$ denotes the supremum with respect to $n\geq{1}$, and the
integration (defining the $n$th moment $K_{n}[{\mathcal{F}}_{2}]$ of
${\mathcal{F}}_{2}$) runs over the $2d$ plane ${\mathcal{P}}$.

\subsection{The nontrivial quasi-self-intersections}

Next, let us introduce the system of the coupled equations which
represent the appropriately regularized Loop equation (\ref{0.9zze}) in the
{\it bootstrap} form suitable for the analysis of the Ansatz (\ref{9.2}). For
this purpose, in the spirit of eq. (\ref{9.6a}), one is to begin with the
decomposition of the matrix element (\ref{1.3zaz}) associated to eq.
(\ref{0.9zze}),
\be
\hat{\mathcal{E}}_{\Lambda}=\sum_{k=1}^{2}
\hat{\mathcal{E}}_{k}~~~~~,~~~~~
<{\bf x}|~\hat{\mathcal{E}}_{k}|{\bf y}>=
<{\bf x}|~\hat{\cal X}_{k}|{\bf y}>
\int\limits_{\Gamma_{xy}\in{{\bf X}_{k}^{D_{k}}}}
{\mathcal{D}}z_{\mu}(t)~e^{-S_{k}(\Gamma_{xy})}~,
\label{9.7}
\ee
where $<{\bf x}|\hat{\cal X}_{k}|{\bf y}>=
{\Lambda}^{4}\cdot{\mathcal{X}}_{k}({\Lambda}{\bf z})|_{{\bf z}=
{\bf x}-{\bf y}}$ are some functions to
be specified later on. Remark that the path-integral representation,
associated to a given ${\cal E}_{k}({\bf z})=
<{\bf z}|\hat{\mathcal{E}}_{k}|0>$, is
determined by certain effective action $S_{k}(\Gamma_{xy})$ together with the
${D_{k}}\leq{4}$ dimensional domain ${\bf X}_{k}^{D_{k}}$
(which the paths $\Gamma_{xy}$ are allowed to fill in). It is also
restricted that $\cup_{k}{\bf X}_{k}^{D_{k}}={\bf X}^{\tilde{D}}$ and
$\sum_{k=1}^{2}\hat{\cal X}_{k}e^{-S_{k}(\Gamma)}=\hat{\cal X}e^{-S(\Gamma)}$
for $\forall{\Gamma}\in{\cap_{k}{\bf X}_{k}^{D_{k}}}$. Furthermore, for the
functions ${\cal E}_{k}({\bf z})$ to implement the proper hierarchy of scales,
one is to impose additionally that ${\cal E}_{2}({\bf z})$ satisfies the
${\cal G}_{2}\rightarrow{\cal E}_{2}$ option of the conditions (\ref{9.8}),
\be
\int d^{4}z~{\cal E}_{2}({\bf z})=
\left({\cal M}/\Lambda\right)^{4}~~~~~~~,~~~~~~~~
<{\bf r}^{2}>_{{\cal E}_{2}}~\sim{~\cal M}^{-2}~.
\label{9.27}
\ee

The simplest way to complete thus defined decomposition (\ref{9.7}) is to
additionally require that ${\bf X}_{1}^{D_{1}}={\bf X}_{2}^{D_{2}}$
so that, utilizing the
Ansatz (\ref{9.2}) and making use of the Leibnitz rule, one transforms the
Loop equation (\ref{0.9zze}) into the pair of the regularized loop
equations\footnote{For the sake of generality, in eq. (\ref{9.12}) we have
retained the place for ${\bf X}^{D_{1}}_{1}\neq{\bf X}^{D_{2}}_{2}$. This
extended version, remaining consistent with eq. (\ref{9.27}), can be deduced
from eq. (\ref{0.9zze}) via a minor generalization of the above arguments
which implies, in particular, that $(-1)^{k}\sum_{p=1}^{2}\varepsilon_{kp}
\hat{\cal X}_{p}e^{-S_{p}(\Gamma)}=\hat{\cal X}e^{-S(\Gamma)}$ for
$\forall{\Gamma}\in{\left({\bf X}^{\tilde{D}}/{\bf X}_{k}^{D_{k}}\right)}$.}
\be
\frac{1}{\lambda}~\hat{\mathcal{L}}<W_{C}>_{k}=
\oint_{C} dx^{\nu}\oint_{C} dy_{\nu}
<{\bf x}|~\hat{\cal X}_{k}|{\bf y}>\times
\label{9.12}
\ee
$$
\times
\int_{{\bf X}^{D_{k}}_{k}} {\mathcal{D}}z_{\mu}(t)~
\tau^{\star}_{k}(C_{xx}|\Gamma_{xy})
<W_{C_{xy}\Gamma_{yx}}>_{k}<W_{C_{yx}\Gamma_{xy}}>_{k}
e^{-S_{k}(\Gamma_{xy})}~,
$$
where we have introduced the functionals (with $\varepsilon_{kl}=-
\epsilon_{lk}$, $\epsilon_{12}=-1$)
\be
\tau^{\star}_{k}(C_{xx}|\Gamma_{xy})=(-1)^{k}\lim_{N\rightarrow{\infty}}~
\sum_{p=1}^{2}\varepsilon_{kp}~
\frac{<W_{C_{xy}\Gamma_{yx}}>_{p}<W_{C_{yx}\Gamma_{xy}}>_{p}}
{<W_{C_{xx}}>_{p}}
\label{9.16}
\ee
which, contrary to the former case (\ref{0.9zaz}),
encode the nontrivial bootstrap between the low- and high-momentum modes.
(As well as in eq. (\ref{0.9zaz}), {\it no} summation over $k$ is implied in
the r.h. side of eq. (\ref{9.12}).)

In conclusion, let us note that the coupling between the $k=1$ and $k=2$ eqs.
(\ref{9.12}) can be handled in a relatively simple way owing to the following
simplification valid in the considered low-energy regime. Indeed, observe
first that the functionals $\tau^{\star}_{k}(..)$ can be formally absorbed by
the redefinition $\bar{S}_{k}(\Gamma_{xy}|C_{xx})=S_{k}(\Gamma_{xy})-
ln\left(\tau^{\star}_{k}(C_{xx}|\Gamma_{xy})\right)$ of the effective action
which brings the pattern of eq. (\ref{9.12}) to the form of eq.
(\ref{0.9zze}). As a result, provided the dependence of the functionals
$\tau^{\star}_{k}(C_{xx}|\Gamma_{xy})$ on $C_{xx}$ and $\Gamma_{xy}$ is
reduced to the dependence on the contours' geometry in the
$1/{\cal M}$-vicinity of $\Gamma_{xy}$ (that is justifiable in the end),
the short- and large-distance equations can be solved {\it separately} for
rather general choice of the latter functionals. Therefore, at the first step,
one can treat these functionals as {\it external sources}, while the proper
matching between the two solutions $<W_{C}>_{1}$ and $<W_{C}>_{2}$ can be
maintained through the pair of the concise conditions deduced from the
requirement that eqs. (\ref{0.9zaz}) and (\ref{9.12}) are consistent for
a given $k$.

\section{The large-distance part}

Concerning the low-energy solution $<W_{C}>_{2}$ of the large-distance $k=2$
equations, the derivation essentially repeats, with a minor modification, the
steps formulated in \cite{Dub4} for the analysis of the $SC$ phase where the
multiscale decomposition is redundant. In compliance with the approach of
\cite{Dub4}, one is motivated to fix the low-energy asymptote $<W_{C}>_{2}$
in the form of the particular large $N$ solution \cite{Dub4} of the $k=2$ eq.
(\ref{0.9zaz}). Implementing the infrared limit of the large $N$
$SC$ expansion \cite{Dub3}, the admissible solution is given by the
(regularized at the scale $\sim{\cal M}$) functional sum
\be
<W_{C}>_{2}~=\sum_{\tilde{M}}~w[\tilde{M}(C)]=
\int\frac{{\cal D}x_{\mu}(\gamma)}{{\cal D}f(\gamma)}~
w[\tilde{M}(C)]
\label{1.6xxx}
\ee
over the genus zero worldsheets $\tilde{M}(C)$ (resulting, for the contours
$C$ without zigzag backtrackings, from the standard immersions
into the Euclidean space ${\bf R^{4}}$ which are conventionally parameterized
by the equivalence classes of the functions $x_{\mu}(\gamma)$) assigned with
the weight $w[\tilde{M}(C)]$ determined by eq. (\ref{0.1}) together with the
identification (\ref{9.9}) constrained by the conditions (\ref{9.8}).
Furthermore, the construction (\ref{1.6xxx})/(\ref{0.1})
is to be complemented by the following prescription \cite{Dub4} to ensure
that $\partial_{\mu}^{{\bf x}(s)}<W_{C}>_{2}=0$, i.e., to implement the
invariance of $<W_{C}>_{2}$ under zigzag backtrackings of the loop
$C$ (which, as a consequence \cite{MigdRep}, allows to make the
translationally invariant Ansatz (\ref{9.2}) consistent with the Bianchi
identities). For this purpose,
starting with any worldsheet $\tilde{M}(C)$ associated to
a {\it nonbacktracking} reference-contour $C\in{\Upsilon'_{nbt}}$, the
particular data of the contour's backtrackings is introduced according to
the concise rule. Being kept intact in the {\it interior}, the worldsheet
$\tilde{M}(C)$ should be deformed on
the boundary $\partial\tilde{M}(C)=C$ so that $C\in{\Upsilon_{nbt}}$ is
mapped onto its backtracking descendant corresponding to the data in question.

Given thus determined solution (\ref{1.6xxx})/(\ref{0.1}) deduced from the
$k=2$ eq. (\ref{0.9zaz}), in subsection 4.1 we then propose the judicious
implementation of the $k=2$ eq. (\ref{9.12}) that, for a generic
quasi-self-intersection of the macroscopic contour $C\in{\Upsilon'}$, makes
it consistent with the latter solution.
As for the conditions sufficient for the existence of such implementation,
akin to the $SC$ phase \cite{Dub4}, one is to require that the considered
solution $<W_{C}>_{2}$, complying with the constraints (\ref{9.27})
together with the correct scaling (\ref{9.18}) of $\sigma_{ph}$, is
infrared equivalent (for macroscopic nonbacktracking contours
$C\in{\Upsilon'}$ of the size
$>>{\cal M}^{-1}$) to the associated large $N$ Nambu-Goto theory
(\ref{1.6xxx})/(\ref{2.5bb}).

Next, to meet the above conditions, in the limit $\lambda\rightarrow{0}$ the
apparent preliminary step is to choose the moments (\ref{0.3csb})
of ${\mathcal{F}}_{2}({\bf z}^{2})$ according to the pattern
\be
K_{0}[{\mathcal{F}}_{2}]=\frac{\eta_{0}}{\lambda}~~~,~~~
K_{2}[{\mathcal{F}}_{2}]=\left(\frac{{\cal M}}{\Lambda}\right)^{4}
\frac{\pi}{2 V_{4}}~~~~~~~,~~~~~~~K_{n}[{\mathcal{F}}_{2}]=
O\left(\frac{\eta_{n}}{\lambda}\right)~~,~~n\neq{0,~2}~,
\label{2.5bxd}
\ee
where $K_{n}[{\mathcal{F}}_{2}]$ denotes the $n$th moment (\ref{0.3csb}) of
${\mathcal{F}}_{2}$, $V_{4}$ stands for the volume of the $4$-dimensional 
ball possessing the unit radius, and we have introduced certain set of
numerical parameters $\eta_{q}$. To specify the admissible domain of these
parameters further, observe first that this domain is constrained, via the
estimate (\ref{0.3csb}) of vortex width $\sqrt{<{\bf r}^{2}>}$, by the second of the
conditions (\ref{9.8}). Complementary, to maintain the ${\cal M}^{2}$-scaling of
(\ref{9.18}) of $\sigma_{ph}$, one is to restrict (see Appendix \ref{macr})
that ${\eta_{0}}/{\lambda}$ is sufficiently larger compared both to unity and
to ${\eta_{n}}/{\lambda}$ for $\forall{n}\geq{1}$.
In particular, according to the general semiclassical
estimate \cite{Luscher} of the characteristic amplitude $\sqrt{<{\bf h^{2}}>}$
of the worldsheet's fluctuations, in thus implemented stringy system
(\ref{1.6xxx})/(\ref{0.1}) the amplitude $\sqrt{<{\bf h^{2}}>}$ is much
larger than the flux-tube's width,
\be
{\mathcal{R}}(s){\cal M}\longrightarrow{\infty}~~~\Longrightarrow{~~~
\frac{<{\bf h^{2}}>}{<{\bf r^{2}}>}\sim
ln[A_{min}(C){\cal M}^{2}]\longrightarrow{\infty}}~,
\label{0.1eea}
\ee
where $A_{min}(C)$ denotes the minimal area of the saddle-point worldsheet
$\tilde{M}_{min}(C)$. Furthermore, (building on the discussion of the
Appendix \ref{macr}) it is presumed that, for macroscopic contours
$C\in{\Upsilon'}$ without zigzag backtrackings, the required infrared
equivalence of the considered solution $<W_{C}>_{2}$ to the properly
associated Nambu-Goto theory (\ref{1.6xxx})/(\ref{2.5bb}) is valid when the
scaling (\ref{9.18}) is maintained.

\subsection{The matching with the $k=2$ eq. (\ref{9.12})}

Now, we are ready to specify the quantities ${\bf X}^{D_{2}}_{2}$ and
$S_{2}(\Gamma_{xy})$ in such a way that ensures the consistency of thus
defined average $<W_{C}>_{2}$ with the $k=2$ loop equation (\ref{9.12}) for a
generic $C\in{\Upsilon'}$. For this purpose, the intermediate step is to
recast the large $N$ stringy pattern (\ref{1.6xxx})/(\ref{0.1}) into the form
written in terms of the combination of the averages entering the r.h. side of
the $k=2$ eq. (\ref{9.12}). It can be done with the help of the following
identity.

\subsubsection{The auxiliary identity to be utilized}

Let the conditions, formulated after eq. (\ref{0.1eea}),
maintain that the stringy system (\ref{1.6xxx})/(\ref{0.1}) (with
${\cal F}_{2}$ chosen as above) is infrared equivalent to the associated
Nambu-Goto theory (\ref{1.6xxx})/(\ref{2.5bb}). Also, for simplicity, we
restrict our attention to the contours $C_{xx}=C_{xy}C_{yx}
\in{\Upsilon'}$ with a single simple nontrivial quasi-self-intersection
when only {\it two} line-segments of $C_{xx}$ quasi-intersect
(i.e. intersect from the low-energy viewpoint) at ${\bf x}(s)={{\bf y}(s')},
~s' \neq{s}$, while the extension to the generic case is straightforward.
The required identity asserts that, in this case,
there is such effective action $S(\Gamma_{xy}|\{{\mathcal{G}}_{p}\})$ that
\be
<W_{C_{xx}}>_{2}=\int\limits_{\Gamma_{xy}\in{\tilde{\bf X}^{3}}}
{\mathcal{D}}z_{\mu}(t)~
\tau^{\star}_{2}(C_{xx}|\Gamma_{xy})
<W_{C_{xy}\Gamma_{yx}}>_{2}<W_{C_{yx}\Gamma_{xy}}>_{2}
e^{-S_{2}(\Gamma_{xy}|\{{\mathcal{G}}_{p}\})}
\label{9.15} 
\ee
where the large $N$ averages $<..>_{2}$ are evaluated
through the considered implementation of the Ansatz
(\ref{1.6xxx})/(\ref{0.1}), and $\tau^{\star}_{k}(C_{xx}|\Gamma_{xy})$ is
defined by eq. (\ref{9.16}). As for the functional integral in eq.
(\ref{9.15}), akin to the $k=2$ eq. (\ref{9.12}), it runs over the paths
$\Gamma_{xy}=\Gamma^{-1}_{yx}\in{\tilde{\bf X}^{3}}$ which are generically
non-self-intersecting in the interior. A given path can be traced back to the connected component
of the cross-section $\tilde{M}(C)\cap \tilde{\bf X}^{3}$ of a particular
genus-zero worldsheet $\tilde{M}(C)$ (entering eq. (\ref{1.6xxx})) with
respect to some $3$-dimensional subspace $\tilde{\bf X}^{3}$ that contains
the selected points ${\bf x}(s)$ and ${\bf y}(s')$. Consequently, the
effective action $S_{2}(..)$ is to be deduced from the stringy sum
(\ref{1.6xxx})/(\ref{0.1}) via the {\it partial} integration over all the
worldsheets $\tilde{M}(C)=\tilde{M}(C_{xy}\Gamma_{yx})\cup
\tilde{M}(C_{yx}\Gamma_{xy})$ which are composed of the two auxiliary surfaces
$\tilde{M}(C_{xy}\Gamma_{yx})$ and $\tilde{M}(C_{yx}\Gamma_{xy})$
spaned by the loops $C_{xy}\Gamma_{yx}$ and $C_{yx}\Gamma_{xy}$ corresponding
to a fixed $\Gamma_{xy}$.

Owing to the imposed constraints, the construction (\ref{9.15})
meets the following two conditions, which are necessary to fulfill
for the consistent utilization of eq. (\ref{9.15}) as the building block
to properly reconstruct the $k=2$ loop equation (\ref{9.12}). On one
hand, the implicit functional dependence of ${S}_{2}(..)$ on
$C_{xy}\Gamma_{yx}$ and $C_{yx}\Gamma_{xy}$, being {\it reduced} to the
dependence on the contours' geometry in the $1/\cal M$ vicinity of the path
$\Gamma_{xy}$, is insensitive to the global geometry of a given macroscopic
contour $C\in{\Upsilon'}$. On the other hand, after the subtraction of a
$\Gamma_{xy}$-independent constant (associated to the contribution accumulated
in the $1/{\cal M}$ vicinity of the points ${\bf x}$ and ${\bf y}$), the
$\Gamma_{xy}$-dependent part of $S_{2}(..)$ is $\Lambda$-{\it independent}
that is crucial for the validity of the second of the constraints
(\ref{9.27}) imposed on $<W_{C}>_{2}$. To justify the former condition, let us
first introduce the modification $\bar{S}_{2}(\Gamma_{xy}|{\mathcal{G}}_{2})
=S_{2}(\Gamma_{xy}|\{{\mathcal{G}}_{p}\})-
ln\left(\tau^{\star}_{2}(C_{xx}|\Gamma_{xy})\right)$ of the
effective action $S_{2}(..)$. The advantage of dealing with $\bar{S}_{2}(..)$
is that the latter action can be interpreted as the relevant measure for the
degree of the {\it deviation} between the weight (\ref{0.1}) and the
'closest' $m_{0}=0$ option of the large $N$ Nambu-Goto pattern (\ref{2.5bb}).
(Indeed, the action $S_{2}$ vanishes when the averages in eq. (\ref{9.15}) are
evaluated through the stringy system (\ref{1.6xxx}) endowed with the latter
option of the Nambu-Goto weight.) In consequence, due to the infrared
equivalence to the Nambu-Goto theory discussed after eq. (\ref{0.1eea}), the
implicit functional dependence of $\bar{S}_{2}(\Gamma_{xy}|{\mathcal{G}}_{2})$
on $C_{xy}\Gamma_{yx}$ and $C_{yx}\Gamma_{xy}$ does exhibit the reduction
required for the original action $S_{2}(..)$. Finally, to justify that this
reduction is indeed extended to $S_{2}(..)$ itself, one is to observe that 
the functional dependence of $\tau^{\star}_{2}(C_{xx}|\Gamma_{xy})$ is
reduced in the same way as well. (The latter property of
$\tau^{\star}_{2}(C_{xx}|\Gamma_{xy})$ is predetermined by
the fact, supported by the analysis of the next section,
that thus implemented decomposition in effect introduces a
finite correlation length $\sim{\cal M}^{-1}$ for the gluonic excitations
associated to the short-distance solution $<W_{C}>_{1}$.) Furthermore, after
the subtraction of certain $\Gamma_{xy}$-independent part, the infrared
asymptote of $ln(\tau^{\star}_{2}(C_{xx}|\Gamma_{xy}))$ is
$\Lambda$-{\it independent} owing to the ${\cal M}$-scaling (\ref{9.30}) of
$m_{1}$. In turn, it allows to fulfill the requirement that the
$\Gamma_{xy}$-dependent part of $S_{2}(..)$ does not depend on $\Lambda$ in
the limit $\Lambda/{\cal M}\rightarrow{\infty}$.

\subsubsection{The final prescription for nontrivial self-intersections}

To complete the construction of such $k=2$ loop equation (\ref{9.12}) that is
consistent (for $C\in{\Upsilon'}$) with the considered solution
(\ref{1.6xxx})/(\ref{0.1}) of the reduced $k=2$ eq. (\ref{0.9zaz}), the final
step is to synthesize the latter reduced equation with the identity
(\ref{9.15}). For simplicity, as previously, we restrict the analysis to
the case when $C\in{\Upsilon'}$ has a single simple quasi-self-intersection.
Then, in the domain of the latter intersection, for any particular pair of
the microscopically separated points ${\bf x}(s),~{\bf y}(s')$ entering the
r.h. side of the $k=2$ eq. (\ref{0.9zaz}), one is to transform
$<W_{C}>_{\infty}$ in compliance with the identity (\ref{9.15}). Furthermore,
one can formally extend\footnote{To this aim, building on eq. (\ref{9.15}),
one is to introduce an analytical continuation of the effective action
$S_{2}(..)$ for an arbitrary pair ${\bf x}(s)$ and ${\bf y}(s')$ including
the case when $s\rightarrow{s'}$.} the latter transformation to any pair of
the points ${\bf x}(s),~{\bf y}(s')$ under the integral in the latter
$k=2$ equation. In sum, under the constraints (\ref{9.27}) combined with the
conditions formulated after eq. (\ref{2.5bxd}),
the Ansatz (\ref{1.6xxx})/(\ref{0.1}) satisfies
the option of the $k=2$ Loop equation (\ref{9.12}) resulting after
the identification $<{\bf z}|\hat{\cal X}_{2}|0>=
<{\bf z}|\hat{\cal G}_{2}|0>$,
${\bf X}^{D_{2}}_{2}=\tilde{\bf X}^{3}$, and $S_{2}(\Gamma_{xy})
={S_{2}(\Gamma_{xy}|\{{\mathcal{G}}_{p}\})}$ (where $\hat{\cal G}_{2}$,
$\tilde{\bf X}^{3}$, and $S_{2}(\Gamma_{xy}|\{{\mathcal{G}}_{p}\})$ are
introduced in eqs. (\ref{9.6a}) and (\ref{9.15}) correspondingly).

At this step, it is also appropriate to clarify the reason why the
consistency of the
multiscale decomposition requires that the physical string tension
$\sigma_{ph}$ can be neither much larger nor much smaller than the squared
$UV$ cut off ${\cal M}$
for the stringy fluctuations. Indeed, as a result of the second of the
constraints (\ref{9.27}) (with the function ${\cal E}_{2}(..)$ defined above),
the subspace
$\tilde{\bf X}^{3}$ in the $k=2$ eq. (\ref{9.12}) can be chosen in the
following way. Namely, once $|{\bf x}(s)-{\bf y}(s')|\leq{\cal M}^{-1}$,
the condition $<L[\Gamma_{xy}]>_{{\cal E}_{2}}\sim{\cal M}^{-1}$ is valid,
where $L[\Gamma_{xy}]$ denotes the length of the path $\Gamma_{xy}$ (entering
eq. (\ref{9.15})), and the averaging $<..>_{{\cal E}_{2}}$ is performed
through the considered implementation of the Ansatz
(\ref{1.6xxx})/(\ref{0.1}). On the other hand, eq. (\ref{9.15}) implies that
$<L[\Gamma_{xy}]>_{{\cal G}_{2}}\sim{1/\sqrt{\sigma_{ph}}}$ once
$|{\bf x}(s)-{\bf y}(s')|\leq{\cal M}^{-1}$. To reconcile the two estimates,
one should impose the ${\cal M}^{2}$-scaling (\ref{9.18}).

\section{The short-distance part}

To reconstruct the remaining short-distance factor $<W_{C}>_{1}$, the idea is
to determine it as the solution, represented through the large $N$ $WC$
series (with the effective $IR$ cut off), to the $k=1$ eq. (\ref{9.12})
extended\footnote{Strictly speaking, this extension requires to continue the
factor $\tau^{\star}_{1}(C_{xx}|\Gamma_{xy})$ to the domain where either
$C_{xy}\Gamma_{yx}$ or $C_{xy}\Gamma_{yx}$ is microscopic so that the Ansatz
can not be directly applied. Nevertheless, in this case, the latter factor
evidently approaches unity while the detailed pattern of the residual
deviation $(\tau^{\star}_{1}(C_{xx}|\Gamma_{xy})-1)$, responsible for the
power-like corrections at short distances, is immaterial for our present
analysis.} to describe both nontrivial and trivial (quasi-)self-intersections
of $C\in{\Upsilon'}$. For this purpose, one is to utilize first that, in terms
of the redefined effective action $\bar{S}_{1}(..)=S_{1}(..)-
ln\left(\tau^{\star}_{1}(..)\right)$, the pattern of the $k=1$ eq.
(\ref{9.12}) matches the pattern (\ref{0.9zze}) of the ordinary Loop equation.
Therefore, by the same token as in \cite{MigdRep}, it is convenient to rewrite
thus implemented $k=1$ equation in the so-called integral form
\be
<W_{C_{xx}}>_{1}=1+\lambda~\hat{\mathcal{L}}^{-1}\times
\left(<W_{C_{xy}\Gamma_{yx}}>_{1}\Bigg|
<W_{C_{yx}\Gamma_{xy}}>_{1} \right)_{\bar{S}_{1}}~,
\label{9.24}
\ee
where the combination $(..|..)_{\bar{S}_{1}}$ symbolically denotes the r.h.
side of the $k=1$ eq. (\ref{9.12}), and $\tau^{\star}_{1}(C_{xx}|\Gamma_{xy})$
can be shown to depend on the contours' geometry only in the
$1/{\cal M}$-vicinity of $\Gamma_{xy}$. As a result, at least in principle,
one is able to iteratively determine $<W_{C}>_{1}$, order by order in
$\lambda$, as the presumably convergent large $N$ weak-coupling expansion
which, in addition to the $UV$ cut off, is endowed with the effective
{\it infrared} cut off at the scale $\sim{\cal M}$. In other words, the
constraints (\ref{9.27}) (together with effect of the
$\tau^{\star}_{1}(C_{xx}|\Gamma_{xy})$-function) are supposed to
ensure that both the gluonic propagator and the vertices, associated to the
latter $WC$ expansion, are nonsingular at zero
momentum. Therefore, one concludes that the loop integrals, characteristic of
the large $N$ $WC$ series in question, can be made free from the spurious
infrared divergences. Altogether, for macroscopic loops $C\in\Upsilon'$, we
expect that the appropriate decomposition supports the perimeter-law pattern
of the low-energy asymptote of
\be                                                      
<W_{C}>_{1}\Big|_{N\rightarrow{\infty}}~\longrightarrow{~
exp\left(-m_{1}L[C]\right)}~~~~~~~~,~~~~~~~~m_{1}=O({\cal M})~,
\label{9.30}
\ee
where, according to the previous section, it is convenient
to maintain the $O({\cal M})$-scaling of $m_{1}$. For this purpose, one
is to employ the judicious implementation (in the spirit of the analytic
regularization \cite{MigdRep}) of the regularized $k=1$ eq. (\ref{9.12}) in
order to suppress those of the $UV$ divergences of $<W_{C}>_{1}$ which
otherwise would result in the unwanted $\Lambda$-scaling of $m_{1}$.

\subsection{The matching with the $k=1$ eq. (\ref{0.9zaz})}

Finally, the $k=1$ eq. (\ref{0.9zaz}) is to be treated as providing with the
simple constraint imposed on the previously determined average $<W_{C}>_{1}$.
To appropriately formulate this constraint, in the limit
$\lambda\rightarrow{0}$ let us first rewrite the latter equation in the form
of the asymptotic relation
\be
\hat{\mathcal{L}}~ln\left(<W_{C}>_{1}\right)=
\frac{\lambda\Lambda^{3}}{\pi}~K_{-1}[{\cal G}_{\Lambda}]\cdot L[C]
\label{9.29}
\ee
where $L[C]$ denotes the length of the perimeter of $C$, while
$K_{-1}[{\cal G}_{\Lambda}]$ is the moment (\ref{0.3csb}) of
${\cal G}_{\Lambda}$ continued to $n=-1$, and we have used that, in the
$WC$ limit (\ref{9.11}), one can neglect the difference between the smearing
functions ${\cal G}_{1}$ and ${\cal G}_{\Lambda}$ entering eq. (\ref{9.6a}).
Next, by-construction, the condition (\ref{9.29}) is to be imposed on
the considered above solution $<W_{C}>_{1}$ only when the macroscopic
contour $C\in{\Upsilon'}$ does not possess nontrivial
quasi-self-intersections. In turn, for such loops, the possibility to
reconcile thus defined average $<W_{C}>_{1}$ with eq. (\ref{9.29}) is
predetermined  by the presence of the effective infrared cut off
built, at the scale $\sim{\cal M}$, into the $k=1$ eq. (\ref{9.12}). Indeed,
the latter cut off ensures that, for $C\in{\Upsilon'}$, in
the limit (\ref{9.11}) the iterative solution $<W_{C}>_{1}$ does fulfill the
asymptotic condition similar to eq. (\ref{9.29}). Therefore, eq.
(\ref{9.29}) can be viewed as the identification of the coefficient
${\lambda\Lambda^{3}}K_{-1}[{\cal G}_{\Lambda}]/{\pi}$ which, being multiplied
by $L[C]$, is equal to the result of the action of $\hat{\mathcal{L}}$ on
$ln\left(<W_{C}>_{1}\right)$. (Furthermore, eq. (\ref{9.30}) implies that
$\hat{\mathcal{L}}~ln\left(<W_{C}>_{1}\right)$ is $\Lambda$-independent in
the weak-coupling limit (\ref{9.11}).)

\section{Conclusions}

We have proposed the low-energy Ansatz (\ref{9.2}) which, being consistent
with the Loop equation, yields the infrared asymptote of the loop average
$<W_{C}>_{\infty}$ in the $D=4$ $U(\infty)\cong{SU(\infty)}$ gauge
theory\footnote{The same strategy can be utilized to construct the low-energy
confining asymptote of the solution associated to the $D=3$ $U(\infty)$ gauge
theory (\ref{1.1}) in the weak-coupling regime of
$(g^{2}N/\Lambda)\rightarrow{0}$.} (\ref{1.1}).
In this particular way, one reconciles the asymptotic freedom (\ref{9.11})
with confinement maintained owing to the stability of the {\it macroscopic}
flux-tubes. The latter vortices, possessing the (transverse) width of order of
the hadronic scale $\sim{\cal M}^{-1}$, are attached to the Wilson loop
source coherently with the point-like gluonic excitations localized at short
distances. For macroscopic loops $C\in{\Upsilon'}$ without zigzag
backtrackings, the flux-tubes are described by the stringy system that, in
the regime (\ref{0.1eea}) of large fluctuations, is infrared equivalent to
the unconventional implementation (\ref{9.18}) of the Nambu-Goto theory
(\ref{1.6xxx})/(\ref{2.5bb}). The simple prescription, to maintain the
mandatory backtracking invariance of $<W_{C}>$, is outlined as well.

To obtain the precise analytic relation between $\sigma_{ph}$ and
$\Lambda_{QCD}$, one will need to specify (through the minimization of the
properly subtracted free energy) the details of how the considered
large $N$ $WC$ series is sewed with the large $N$ $SC$ expansion
at the scale which is supposed to be of order of ${\cal M}^{-1}$.
For this purpose, it is necessary to handle the harder case of intermediate
and small, in the units of ${\cal M}^{-1}$, contours that requires to resolve
the full-fledged bootstrap between the eqs.
(\ref{9.12}). Nevertheless, even without this
extension, the present analysis predicts the qualitatively admissible
${\cal M}^{2}$-scaling (\ref{9.18}) of $\sigma_{ph}$ which entails the
important implications.

In consequence of eq. (\ref{9.18}), to facilitate the quantum
analysis in the regime (\ref{0.1eea}), the infrared asymptote of the stringy
sum (\ref{1.6xxx})/(\ref{2.5bb}) is to be reformulated
(in the spirit of the Pauli-Villars regularization)
as the somewhat unconventional stringy theory. Possessing the $UV$
cut off $\bar{\Lambda}$ with $\sqrt{\sigma_{ph}}/\bar{\Lambda}\rightarrow{0}$,
{\it this theory should not exhibit propagating degrees of freedom at the
short-distance scales $<<{1/\sqrt{\sigma_{ph}}}$, while approaching
the proposed Nambu-Goto pattern (\ref{1.6xxx})/(\ref{2.5bb})
in the infrared domain at the scales sufficiently larger than
$1/\sqrt{\sigma_{ph}}$}. To formalize this feature, one has to
reformulate the infrared asymptote of the nonrenormalizable (from the
power-counting viewpoint) stringy system (\ref{1.6xxx})/(\ref{2.5bb}) in
terms of certain {\it renormalizable} noncritical string which, in the
spirit of the conventional Polyakov's approach, should include the
Liouville field possibly together with some other auxiliary fields on the
worldsheet. In particular, the constraint (\ref{9.18}) translates into the
condition that the required renormalizable noncritical theory must be
devoid of any $UV$ divergences. A plausible candidate for the '$UV$ finite'
model, complying with the above requirements, may be the
specific dilatonic extension \cite{Burw} (considered on the flat background
$\hat{R}=0$) of the Liouville noncritial string with the properly identified
parameters. The promising sign is that the tentative stringy system
\cite{Burw}, possessing the critical dimension $D_{cr}=24$, does {\it not}
exhibit imaginary critical coefficients in $D=4$. In turn, it presumably
foreshadows that the string, dual to the gauge theory, does avoid the
notorious branched polymer instability.

\begin{center}
{\bf Acknowledgements.}
\end{center}

This project is partially supported by grants CRDF RP1-2108 and INTASS-390.

\app{The $IR$ equivalence to the Nambu-Goto theory}
\label{macr}

Let us deduce the conditions when the stringy system
(\ref{1.6xxx})/(\ref{0.1}) approaches its low-energy asymptote in the form of
the unconventional implementation (\ref{9.18}) of the Nambu-Goto theory
(\ref{1.6xxx})/(\ref{2.5bb}). For a preliminary orientation, observe first
that, under certain conditions, the quasi-local weight (\ref{0.1}) can
be interpreted as a sort of smearing (or regularization) of the Nambu-Goto
pattern (\ref{2.5bb}) local on the worldsheet $\tilde{M}$. To support this
interpretation, consider first such deformation of the pattern (\ref{0.1})
that implements the formal limit when the flux-tube's width (\ref{0.3csb})
vanishes. The required deformation is obtained (neglecting the conditions
(\ref{9.8})) when the smearing function is {\it analytically
continued}, ${\cal M}^{2}{\cal F}_{2}({\cal M}^{2}({\bf x}-{\bf y})^{2})
\rightarrow{K_{0}[{\cal F}_{2}]\delta^{w}_{2}({\bf x}-{\bf y})}$ (where
$K_{0}[{\cal F}_{2}]$ stands for the zeroth moment (\ref{0.3csb}) of
${\cal F}_{2}$), to become proportional to the 2-dimensional delta-function
$\delta^{w}_{2}({\bf x}(\gamma)-{\bf y}(\gamma'))$ on the worldsheet
$\tilde{M}$, with ${\bf x},{\bf y}\in{\tilde{M}}$. In this limit, the stringy
sum (\ref{1.6xxx}) is known to be
invariant (see \cite{LE/MM,Dub4}) under the substitution
of thus deformed weight (\ref{0.1}) by the $m_{0}=0$ option of the Nambu-Goto
weight (\ref{2.5bb}) with the bare string tension $\sigma_{0}=\lambda
K_{0}[{\cal F}_{2}]{\cal M}^{2}/2$.

As a result, the asserted infrared equivalence can be
heuristically supported by the following argument.
Reinterpreting the regime (\ref{0.1eea}) as the low-energy limit of the
stringy system (\ref{1.6xxx})/(\ref{0.1}) applied to macroscopic loops,
one may expect that thus introduced regularization of the Nambu-Goto system
becomes infrared unobservable. To make these arguments precise, one is to
adapt the general approach of the {\it Wilsonean} renormgroup that is the
subject of a forthcoming paper. Here, we restrict the analysis to the
discussion of the following large $\eta_{0}$ asymptote (where $\eta_{n}$
is introduced in eq. (\ref{2.5bxd})) of $<W_{C}>_{2}$ which strengthens the
above heuristic argumentation. Complementary, it supports that both the
scaling (\ref{9.18}) and the infrared equivalence are supposed to be
maintained once ${\eta_{0}}/{\lambda}$ is sufficiently larger compared both to
unity and to ${\eta_{n}}/{\lambda}$ for $\forall{n}\geq{1}$.

To deduce the auxiliary asymptote in question, note first that, in view of
the estimate (\ref{0.3csb}), the smearing function
${\cal F}_{2}$ approaches its zero-width deformation (sketched above)
in the specific formal limit. The limit is defined by the requirement that,
keeping $\lambda$ small but finite, ${\eta_{0}}$ is sent to infinity, while
the remaining parameters ${\eta_{n}}$ are chosen in such a way that the
corresponding vortex width (\ref{0.3csb}) scales as
${\cal M}^{-1}/\eta_{0}$ which, in particular, ensures the condition
(\ref{9.18}). In this limit, for any given macroscopic contour $C$,
the characteristic amplitude of the flux-tube's fluctuations
$\sim{\cal M}^{-1}\times
\left(ln[A_{min}(C){\cal M}^{2}]/\eta_{0}\right)^{1/2}$
is much larger than the vortex width $\sim{\cal M}^{-1}/\eta_{0}$.
Consequently, the leading $\eta_{0}\rightarrow{\infty}$
semiclassical asymptote of thus deformed quasi-local Ansatz results in
\be
<W_{C}>_{2}~\longrightarrow{~
exp\left(-\frac{\eta_{0}{\cal M}^{2}}{2}A[\tilde{M}_{min}(C)]\right)}~,
\label{1.6dds}
\ee
where $A[\tilde{M}_{min}(C)]$ is the minimal area of the saddle-point surface
$\tilde{M}_{min}(C)$ spanned by
a macroscopic contour $C$ belonging to the subspace ${\Upsilon'}$
(of the full loop space $\Upsilon$) appropriately chosen in the end of this
Appendix. In turn, in the considered limit, the infrared equivalence in
question is justified by the straightforward semiclassical computation which
ensures that the same asymptote (\ref{1.6dds}) takes place in the large $N$
Nambu-Goto theory (\ref{1.6xxx})/(\ref{2.5bb}) after the identification
$\bar{\lambda}=\eta_{0}$.

To complete the discussion of the considered infrared equivalence, let us
make precise the twofold requirement that selects the appropriate subspace
$\Upsilon'$ utilized throughout the paper. To begin with, evidently one is to
impose that the radius ${\mathcal{R}}(s)$ of the contour's curvature is
$>>{\cal M}^{-1}$. Taking into account eq. (\ref{0.1eea}), it can be shown to
entail the weaker variant of the required equivalence leaving space for
possible deviation in the pattern of the boundary term which, e.g., in the
presence of the macroscopic one-dimensional self-intersections of $C$,
exhibits more complicated pattern compared to the second term in the exponent
of (\ref{2.5bb}).
To avoid this deviation (which otherwise would hamper the consistency of
the system (\ref{1.6xxx})/(\ref{0.1}) with the Loop equation), we have to
introduce the additional restriction on the geometry of $C$. Apart from the
$1/\cal M$-vicinity of a finite number of
points (of quasi-self-intersection), the distance
$|{\bf x}(s)-{\bf y}(s')|$ should be much larger than the flux-tube's width
$\sqrt{<{\bf r^{2}}>}\sim{{\cal M}^{-1}}$,
provided the length $L_{xy}$ of the corresponding segment of the
boundary contour $C=\partial \tilde{M}$ is much larger than ${\cal M}^{-1}$.
In short, the latter definition excludes those geometries of $C$ which,
from the macroscopic viewpoint, can be interpreted as one-dimensional
self-intersections.

\enddocument